# Spin-polarized electronic structure of the core-shell ZnO/ZnO:Mn nanowires probed by x-ray absorption and emission spectroscopy†

A A Guda*[a], N Smolentsev[a], M Rovezzi[b], E M Kaidashev[c], V E Kaydashev[c], A N Kravtsova[a], V L Mazalova[a], A P Chaynikov[a], E Weschke[d], P Glatzel[b] and A V Soldatov[a]

[a] *Research center for nanoscale structure of matter, Southern Federal University, Zorge 5, 344090, Rostov-on-Don, Russia. Tel:+7-(863)-2975-326; E-mail:guda_sasha@mail.ru, guda@sfedu.ru*
[b] *European Synchrotron Radiation Facility, 6 rue Jules Horowitz, BP220, 38043 Grenoble, France*
[c] *Vorovich Research Institute of Mechanics and Applied Mathematics, Southern Federal University, pr. Stachki 200/1, 344090, Rostov-on-Don, Russia*
[d] *Helmholtz-Zentrum Berlin für Materialien und Energie, Wilhelm-Conrad-Röntgen-Campus BESSY II, Albert-Einstein-Str.15, D-12489, Berlin, Germany*

The combination of x-ray spectroscopy methods complemented with theoretical analysis unravels the coexistence of paramagnetic and antiferromagnetic phases in the $Zn_{0.9}Mn_{0.1}O$ shell deposited onto array of wurtzite ZnO nanowires. The shell is crystalline with orientation toward the ZnO growth axis, as demonstrated by X-ray linear dichroism. EXAFS analysis confirmed that more than 90% of Mn atoms substituted Zn in the shell while fraction of secondary phases was below 10%. The value of manganese spin magnetic moment was estimated from the Mn Kβ X-ray emission spectroscopy to be $4.3\mu_B$ which is close to the theoretical value for substitutional $Mn_{Zn}$. However the analysis of $L_{2,3}$ x-ray magnetic circular dichroism data showed paramagnetic behaviour with saturated spin magnetic moment value of $1.95\mu_B$ as determined directly from the spin sum rule. After quantitative analysis employing atomic multiplet simulations such difference was explained by a coexistence of paramagnetic phase and local antiferromagnetic coupling of Mn magnetic moments. Finally, spin-polarized electron density of states was probed by the spin-resolved Mn K-edge XANES spectroscopy and consequently analyzed by band structure calculations.

## 1 Introduction

Being potential building blocks for spintronics manganese-doped ZnO nanostructures attract significant attention of experimental and theoretical groups[1]. Manganese dopants in the ZnO host lattice act as deep donors and tune its magnetic properties and conductivity[2]. The room-temperature ferromagnetism (RTFM) was theoretically predicted for diluted ZnO:Mn[3] and experimentally observed shortly thereafter[4]. However, the origin of RTFM is still under debate. The measurements of magnetization for ZnO:Mn performed using superconducting quantum interference device (SQUID) reveal its anisotropic behaviour[5], strain-dependency[6], switching from paramagnetic to ferromagnetic behaviour when the temperature increases from 5 to 300K[7] or even complete absence of magnetization under certain conditions [8]. It is the lattice distortions around transition metal (TM)[9] or additional structural defects that play a key role in the observed FM properties rather than TM itself[8, 10]. This argument is supported by the observation of the ferromagnetism even in the pure ZnO[11-13] nanostructures. Thus there is a growing consensus that the high-temperature ferromagnetism reported for ZnO:TM is related either to the contamination during the sample manipulation or to secondary phases[14-16].

Combining SQUID and x-ray magnetic circular dichroism (XMCD) measurements Ney *et al.*[17] found a coexistence of paramagnetic behaviour of Co ions and their antiferromagnetic (AFM) coupling in ZnO:Co thin films. Using similar technique it was also discovered[18] that RTFM is not related to the Mn magnetic moments in ZnO since they reveal a pure paramagnetic behaviour. AFM coupling between TM diluted in ZnO could be probed by XMCD measurements with high magnetic field up to 17T that allow determining accurate M(H) curve[19]. However, when N co-doping was introduced, 1% of ferromagnetic phase was found in ZnO:Mn thin films[20].

In the present paper we determine a concentration of the AFM phase in the diluted ZnO:Mn shell of core-shell ZnO/ZnO:Mn nanowires using advanced methods of x-ray absorption spectroscopy (XAS), x-ray emission spectroscopy (XES) and theoretical analysis employing multiplets and density functional theory. We report the first measurements of the Mn K-edge XANES in ZnO:Mn collected separately for different spin projections with respect to the d-shell magnetic moment making use of the spin-selectivity of XES[21-23].

The value of Mn magnetic moment was obtained separately by analyzing the chemical shift of the Mn Kβ XES spectra and from theoretical analysis of Mn $L_{2,3}$ XMCD spectra measured at moderate fields up to 5T. Quantitative analysis of XMCD data was performed. It was found that AFM coupling of Mn magnetic moments diluted in ZnO can explain the lower value of spin magnetic moment obtained from XMCD. Analysis of XAS spectra using non-muffin-tin approach and extended X-ray absorption fine structure (EXAFS) data support this observation.

## 2 Experiment

ZnO nanowires were grown by pulsed laser deposition (PLD) in argon flow (50 sccm) at high pressure (100 mbar) using Au catalyst-assisted vapour-liquid-solid (VLS) mechanism as described elsewhere[24]. The temperature of the substrate was maintained at 830 ºC. The shell made up of $Zn_{0.9}Mn_{0.1}O$ film was deposited by 2000 laser pulses employing the off-axis PLD method at 0.2 mbar oxygen pressure and a substrate temperature of 550 ºC as reported previously[25].

Mn $L_{2,3}$ X-ray XMCD spectra were measured in the undulator UE46/1-PGM-1 beamline of BESSY-II, Helmholtz-Zentrum, Berlin. The sample was placed in an ultra-high vacuum chamber with a magnetic field perpendicular to the sample surface. The temperature of the sample holder was maintained at 10 K. The total electron yield from the sample surface was collected for different X-ray energies and circular polarizations.

X-ray emission and high energy resolution fluorescence detection (HERFD) XANES/EXAFS spectra were collected at the ID26 beam-line [26, 27] of the European Synchrotron Radiation Facility in Grenoble, France. Si(311) double-crystal monochromator was employed to scan the incident energy and five Si(110) analyzer crystals working at (440) reflection were used for fluorescence radiation. For the HERFD XANES the incident energy was varied around Mn K absorption edge in the range of 6530:6620 eV while the fluorescence detection energy was fixed subsequently to $K\beta_{1,3}$ and $K\beta'$ emission lines with energies at 6491.8 eV and 6476.2 eV, respectively.

## 3 Computational details

Electronic structure of the Mn-doped ZnO and Mn K-edge XANES spectra was simulated by means of full-potential linearized augmented plane-wave approximation (FLAPW) implemented in the Wien2k program package[28, 29]. The generalized gradient approximation within Perdew, Burke, Ernzerhof exchange-correlation functional (GGA PBE)[30] was used. The following initial parameters of the wurtzite ZnO were used: a=b=3.25Å, c=5.21Å, space group P63mc and atomic coordinates Zn(1/3, 2/3, 0), O(1/3, 2/3, 0.382) [31]. The influence of the supercell and basis set size was explored (see supplementary information). It was found that even a 2x2x1 supercell is sufficient to reproduce the experimental Mn K-edge XANES well. All geometries were relaxed until atomic forces were less than 0.03eV/Å. Energy convergence criterion in the self-consistent iteration procedure was set to 1meV. For the Mn $L_{2,3}$ XANES and XMCD spectra atomic multiplet simulations[32] were performed, as implemented in CMT4XAS program[32]. The computational parameters applied will be discussed further in the paper.

## 4 Results and discussion

### 4.1 Structure analysis.

Wurtzite ZnO/ZnO:Mn nanowires have a rod shape and grow preferentially perpendicular to the substrate[33]. Energy filtered transmission electron microscopy shows that Mn is distributed preferentially in the shell while the core is pure ZnO. The samples show a high exciton to green band intensity ratio even at room temperature[25]. As confirmed by X-ray diffraction[34] high-pressure PLD process used in the experiment produces high-quality Mn doping of ZnO, free of secondary phases (see Figure S3 in supplementary information). However, standard laboratory X-ray diffraction fails to detect small amounts of dopant-related secondary phases[35] and nanosized precipitations do not produce pronounced peaks in the diffraction pattern. For this reason, the Mn local atomic structure was investigated by Mn K-edge x-ray absorption fine structure.

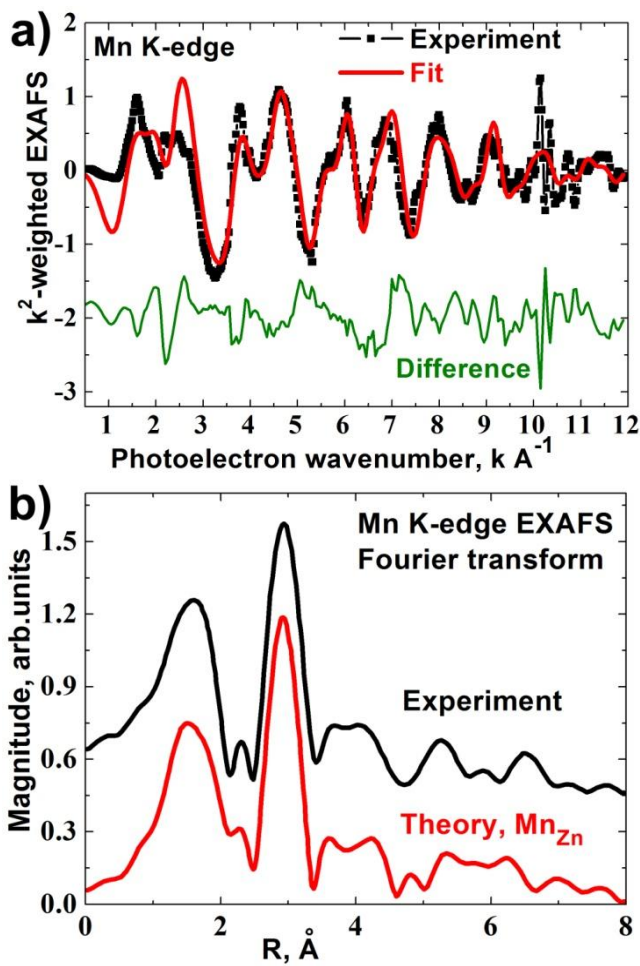

**Figure 1**. EXAFS data analysis. (a) $k^2$-weighted EXAFS data (black squares) is compared to the simulation of 90% $Mn_{Zn}$ substitutional defects in ZnO (red line). The bottom curve shows the difference between experiment and simulation. (b) The Fourier transform of the EXAFS data for experimental data and simulations for $Mn_{Zn}$ model.

Figure 1 shows the results of the EXAFS analysis at Mn K-edge. The experimental data are compared to the multiple scattering[36] simulation of a Mn substitutional defect in ZnO ($Mn_{Zn}$). The experimental thermal and structural disorder damping was reproduced by using, respectively, a correlated Debye model with $\Theta_D$ = 400 K and adding a global $\sigma^2$ = 0.005. As expected, a 3% expansion of the Mn-O first shell distances with respect to the bulk Zn-O was found. To reproduce the overall amplitude of the experimental data, the simulated spectrum has been least-squares fitted with an amplitude parameter. We considered several possible positions of Mn atom in the ZnO lattice and secondary phases: Mn in the Zn site, octahedral and tetrahedral interstitials, metallic Mn, Mn oxides (MnO, $Mn_2O_3$, $Mn_3O_4$, $MnO_2$). When linear fit was performed using two components ($Mn_{Zn}$ and other model) the concentration of $Mn_{Zn}$ phase was found to be in the range 90…95%. The fitting procedure using combination of three components gave similar results. Thus we conclude that 90±5% of Mn substitute Zn. Additional details of the fitting procedure can be found in the supplementary information of Ref[37]. The level of noise which is present in the difference spectrum of Figure 1a makes it difficult to identify the remaining fraction. The latter can be attributed mainly to minor octahedral coordinated phases such as Mn-oxides since the formation energy of interstitial defects is high. Independent analysis of the near-edge energy region of spectra for ZnO:Mn and Mn oxides measured with high energy resolution (Figure 2, Figure 4a and Figure S4 in the supplementary information) supports the conclusion on low concentration of secondary phases.

Figure 2 shows the Mn K-edge XANES and XLD spectra. Complemented by theoretical analysis XANES provides information about structure distortions around Mn dopants inside ZnO[9, 38-40] which is complementary to the EXAFS data analysis. Theoretical simulations for the substitution $Mn_{Zn}$ defect reproduce the shape of both XANES and XLD. Thus, ZnO:Mn shell deposited on the vertically aligned ZnO nanowires is not amorphous but crystalline and has a preferential direction of crystal growth parallel to the orientation of the nanowires. The calculated XLD signal is twice as large as experimentally observed. This can be interpreted by certain disorientation of crystallites in the ZnO:Mn shell as well as the presence of non-vertical ZnO nanowires. The XANES spectra measured with different linear polarizations are useful to detect oxygen vacancies[41] and provide a very accurate method for determining the phase purity in non-cubic single crystals as was demonstrated in case of ZnCoO thin films[14]. On the other hand, it loses sensitivity when dealing with polycrystalline or cubic phases. Conversely, XAS is sensitive to the probed element (Mn in our case), regardless the degree of crystallinity, orientation and symmetry. In the present case, where the nanowires are covered with a $Zn_{0.9}Mn_{0.1}O$ shell, combining XAS and XLD permits one to state that the reduced XLD signal (~50% of the theoretical one) is due to the fact that part of the shell is not coherently oriented to the substrate.

It is well-known that no cubic symmetry gives rise to linear dichroism in dipole transitions[42]. The inset in Figure 4b shows that pre-peak A originates from $O_p$-$Mn_d$ hybridized states localized mainly in the $MnO_4$ tetrahedron. Therefore the pre-edge peak A in Figure 2b does not show linear dichroism since local order around $Mn_{Zn}$ in wurtzite ZnO is close to the $T_d$ symmetry similarly to the zinc blende structure. $T_d$ symmetry breaks down only in the second coordination sphere of wurtzite structure (in Figure 2a the upper tetrahedron in the wurtzite structure is rotated by 60 degrees as compared to zinc blende). Thus higher energy region of XANES reveals dichroism because it is characterized by the large mean free path of the photoelectron[43].

The energy region above the absorption edge marked as C in Figure 2b is sensitive to the local disorder around TM in ZnO, *e.g.* to the presence of oxygen vacancies[40, 41, 44]. Several groups have considered different types of structural defects in order to explain the experimental spectrum of ZnO:Mn[9, 40, 45]. These works rely on the full multiple scattering formalism within the muffin-tin approximation (MTA). The shape of the spectrum calculated for $Mn_{Zn}$ defect using the full-potential band structure simulations [46] differs from those obtained with MTA[46]. In Figure S2 of supplementary information we compare the Mn K-edge spectra calculated using the same structural parameters and energy broadening following the MTA cluster approach and full-potential band structure method. Two approximations yield similar spectral shape of Mn K-edge, but with different intensities of the peaks $C_2$,$C_3$ (see theoretical black and blue curves in Figure S2). MTA overestimates this intensity and one should consider this effect when structural models are analyzed.

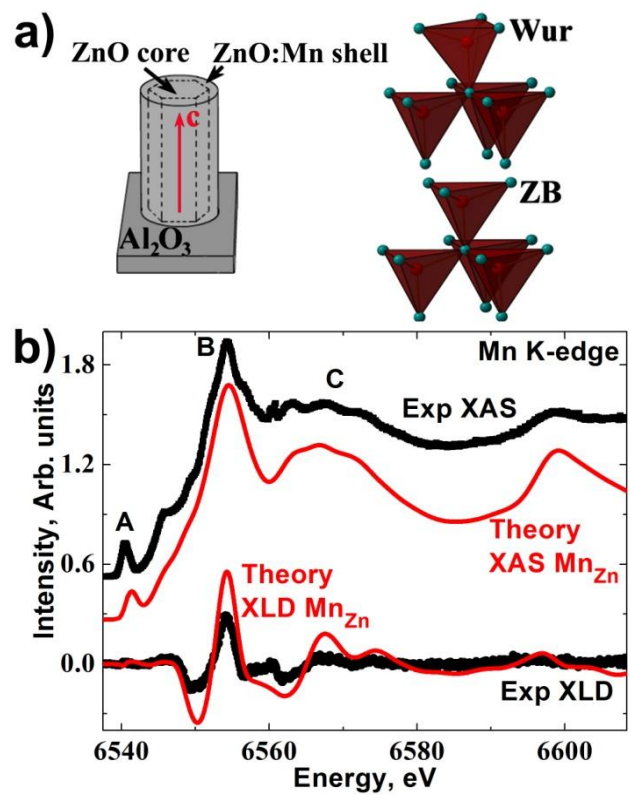

**Figure 2**. (a) the schematic view of the nanowire morphology and local atomic structure around substitutional $Mn_{Zn}$ (b) Experimental Mn K-edge XANES and XLD compared to the simulations for substitutional $Mn_{Zn}$.

### 4.2 Absolute values of spin and charge of Mn atom

Mn Kβ x-ray emission spectra result from 3p - 1s transitions and consist of the main line, $Kβ_{1,3}$, and a satellite, Kβ'. A single electron scheme of these transitions is shown in Figure 3a. The spin orientation of the 3p shell and 3d shell can be parallel or antiparallel in the final state, when 1s core hole is filled. These two states have different energies and therefore two emission lines are observed in the spectrum. Strong final-state 3p3d exchange coupling results in a sensitivity of XES to the 3d population and to the relative spin orientation of the 3p and 3d electrons[22].

X-ray emission spectra for the ZnO/ZnO:Mn nanoneedles and reference compounds are shown in Figure 3b. Samples were irradiated by the non-resonant excitation at 6700 eV. The spectra exhibit a chemical shift, *i.e.* the energy position of the $K\beta_{1,3}$ line depends on the Mn spin state and therefore its charge state. More precisely the energy separation between the main line $K\beta_{1,3}$ and the satellite $K\beta'$ ($\Delta E$) is proportional[47] to the Slater exchange integral (J) between the 3p and 3d electrons, and to the number (2S) of unpaired electrons in the 3d shell: $\Delta E = J(2S+1)$. This approximation has proved to be a valid tool for analyzing experimental data[48,49] and one can calculate absolute values of the integrated difference spectra (IAD) with respect to a certain compound in order to follow the evolution of $\Delta S$ quantitatively. We took a $MnO_2$ XES as a reference with smallest spin magnetic moment. Then the IAD for $Mn_2O_3$, MnO, ZnO:Mn relative to $MnO_2$ were calculated. Figure 3c shows the IAD values for Mn-oxide reference compounds plotted as a function of the Mn spin magnetic moment S. Magnetic moments were calculated around Mn atoms in the sphere which was used in GGA-PBE FLAPW approximation. By assuming the linear evolution of S for the oxides[50] the value of S for ZnO:Mn was then taken from this fit. The value found is $S=4.3\mu_B$. This value is lower than $5\mu_B$ in case of ideal $d^5$ configuration because the hybridization between Mn d- and oxygen p-states results in a magnetization of the four oxygen atoms in the first coordination sphere around Mn as shown in the inset of Figure 3c. XAS data also show a visible chemical shift of the main edge for the series of Mn oxides and ZnO:Mn, although the shifts in the XES are more linearly correlated with the oxidation state and depend less on the atomic configuration[51].

Using the spin-sensitivity of x-ray emission spectra we have measured spin-selective x-ray absorption spectra[21-23] as shown in Figure 4a. Ligand-field multiplet theory demonstrates that the $K\beta'$ emission line arise from 100% spin-up transitions, while the $K\beta_{1,3}$ is mostly spin-down (see Figure 6 in Ref.[52]); this is schematically shown using one-electron approximation, Figure 3a. Thus the spin-selective x-ray-absorption spectra were obtained by fixing emission energy at the certain energies of the satellite $K\beta'$ emission line (6476.2 eV) and main $K\beta_{1,3}$ emission line (6491.8 eV) respectively, while scanning the excitation energy through the Mn K absorption edge. The difference between spin-up and spin-down spin-selective XANES is in direct ratio with the difference in the spin polarization of the empty states multiplied by their matrix element. The common hard x-ray probe of spin magnetic moment, Mn K-edge XMCD, is proportional to this value with the energy dependent proportionality factor, so-called Fano factor. The latter can be determined either experimentally[53] or theoretically in order to extract the degree of polarization and spin magnetic moment value from the K-edge XMCD spectra. In contrast to XMCD spin-selective XANES spectra do not require circular polarization of photons. Coherent orientation of magnetic moments is also not necessary because spin-sensitivity arises from internal localized manganese 3p-3d exchange interaction.

Experimental spectra for different spin polarizations reveal different fine structure. In Figure 4a the main maximum $B_2$ in spin-minority spectrum is shifted to the higher energy compared to the spin-majority spectrum due to exchange interactions of the photoelectron and Mn magnetic moment. The localized unoccupied electronic states correspond to the Mn K-edge pre-peak A and are totally spin-polarized. Conversely conduction states (shoulder $B_1$) are partially spin-polarized as follows from the different intensities of spin-minority and spin-majority XANES. These results are consistent with theoretical simulations of the spin-polarized electron density of states. Unoccupied manganese d-projected DOS both in ground state (Figure S5 in supplementary information) and when 1s core hole is created

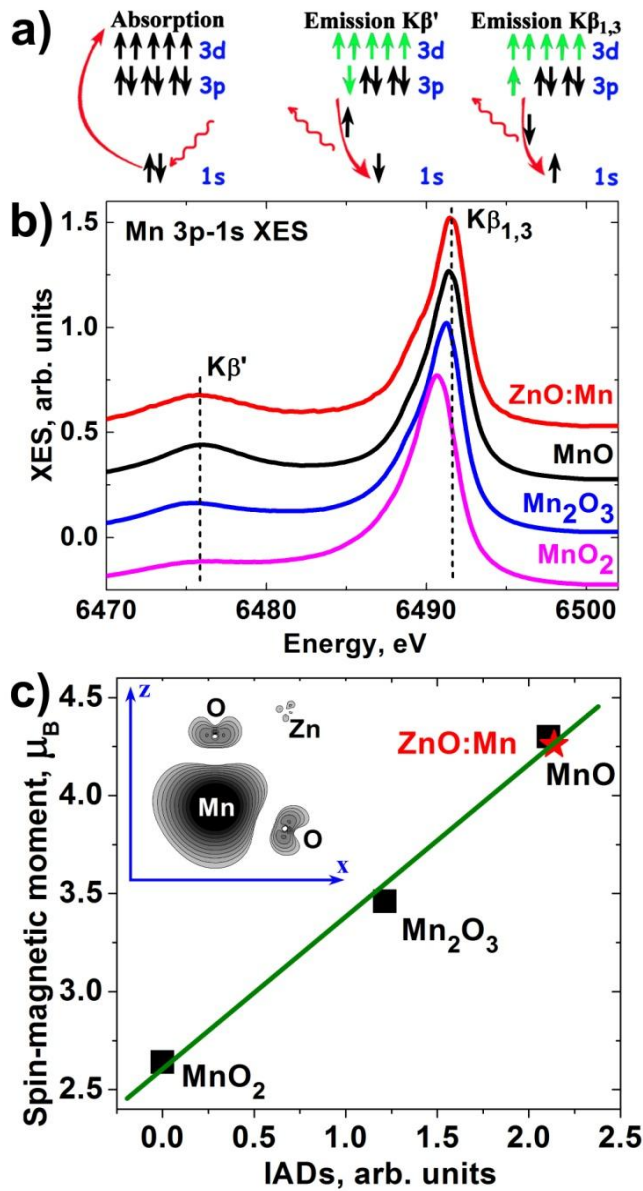

**Figure 3**. XES data analysis. (a) Single electron scheme of the transitions corresponding $K\beta_{1,3}$ and $K\beta'$ lines. (b) Experimental $K\beta$ XES for the series of Mn oxides and Mn doped ZnO core-shell nanowires. Dashed lines indicate maxima of $K\beta'$ (6575.6 eV) and $K\beta_{1,3}$ (6491.7 eV) lines of ZnO:Mn. (c) Integrated absolute values of the difference XES spectra of Mn-oxides with respect to $MnO_2$. The linear dependence of the Mn-oxides' IADs is fitted (green line) and the data point of ZnO:Mn is then placed onto this line. The inset shows the spin magnetic moment distribution around $Mn_{Zn}$ defect inside ZnO, z-axis is aligned with c-axis of wurtzite.

(Figure 4b) are totally spin-polarized. The simulations were performed for the ZnO supercell containing one Mn defect, which corresponds to the total ferromagnetic order. In case of paramagnetic or antiferromagnetic materials our discussion about spin-polarization is then referred to the local orientation of the d-shell magnetic moment of the absorbing atom.

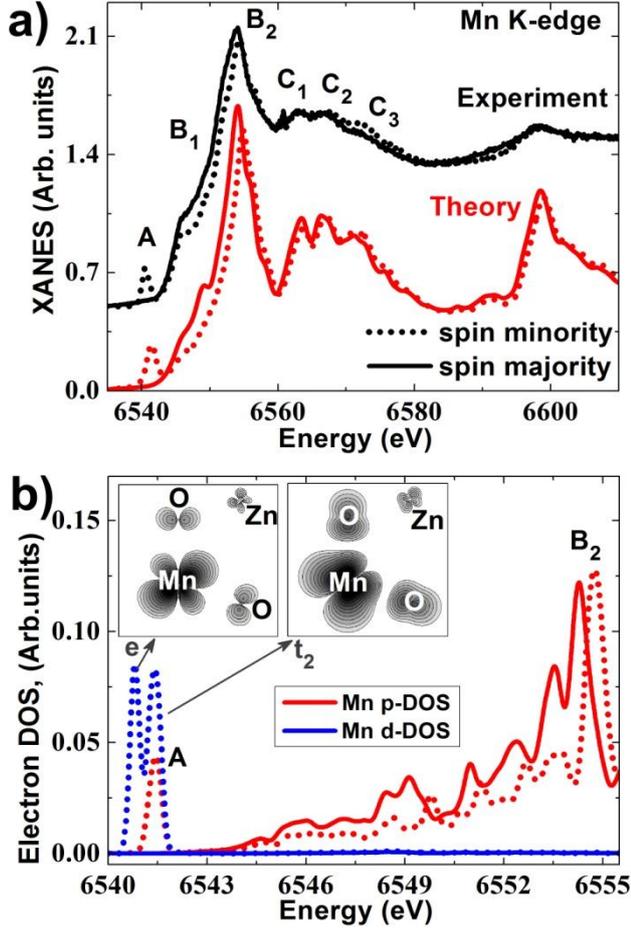

**Figure 4.** (a) Experimental Mn K-edge XANES measured for different spin polarizations compared to the FLAPW simulations for the $Mn_{Zn}$ substitution defect (please note that no standard Mn K-edge XMCD were measured – see discussion in the text). Solid curves stand for spin majority, dotter curves – for spin minority. (b) Calculated Mn p-projected (red) and d-projected (blue) unoccupied electron DOS for the $Mn_{Zn}$ defect with the core hole on Mn 1s. Solid lines stand for spin majority, dotted lines – for spin minority. DOS are shifted energetically to the same energies as XANES. Mn-d states are multiplied by the factor 0.02. Insets show electron density log-scale distribution for the energy intervals that correspond to E and $T_2$ states.

Experimental spectra are complemented by the theoretical simulations (red curves in Figure 4a). Energy broadening for the theoretical spectra is intentionally smaller than in Figure 2 in order to highlight spectral features. The theory agrees well with the measured data and confirms that spin-polarized unoccupied electron DOS is observed in the experiment. Subsequently the same theoretical approximation is used for the electronic structure analysis. Figure 4b shows the calculated spin-polarized density of unoccupied electron states for a supercell with $Mn_{Zn}$ defect in the presence of a Mn 1s core hole. Crystal field splits Mn d states into e and $t_2$ manifolds as shown in the inset of Figure 4b. Peak A observed in the experimental data corresponds to the hybridized $Mn_d$-$O_p$ states of $t_2$ symmetry and these oxygen p-states were observed in the oxygen K-edge XANES of $Zn_{1-x}Mn_xO$ films[54]. Group theory restricts hybridization and therefore electrical dipole transitions from Mn 1s only to the $t_2$ states. Thus Mn p-DOS (red curve) that is observed in the dipole s→p transition is equal to zero beneath peak e of Mn d-DOS (blue curve).

The energy position of the localized electronic states in the pre-edge peak A in Figure 4a relative to the delocalized states in the absorption maximum $B_2$ can be used as a quantitative measure for the quality of theoretical electron-correlation approximations. Peak A originates from the localized Mn d-states that are generally not described well in the standard GGA approximation so one might think of applying self-interaction corrections within orbital-dependant GGA+U to improve the results[55, 56]. Such corrections increase the band gap energy, shift occupied states to the lower energies and unoccupied ones to higher energies[57]. We have simulated Mn K-edge XANES for the substitutional $Mn_{Zn}$ also in GGA+U approximation by applying $U_{eff}$=8.5eV for Zn[58] d-states and $U_{eff}$=4.5eV for Mn. The following energy difference Δ between A and $B_2$ peaks of the Mn K-edge were obtained: $\Delta_{GGA}$=13.33eV, $\Delta_{GGA+U}$=12.95 eV. Supplementary information (see Figure S6) contains series of calculated spectra for different $U_{eff}$ values that show a linear decrease of Δ when $U_{eff}$ increases. Experimental value $\Delta_{EXP}$=13.77 eV is closer to the GGA results. Thus a Hubbard correction does not improve the GGA results for the energy position of the unoccupied Mn d-states although it is considered to be necessary for the occupied TM d-states[59].

The possible explanation of such discrepancy can be related to the core-hole effects. Present simulations rely on the one-electron final state approximation when XANES spectrum is calculated according to the Fermi golden rule. The final states are calculated self-consistently in the presence of a core hole in the Mn 1s level. Deviations from the Final state rule due to relaxation of the electron system can influence the core-hole screening. This, in turn, changes the energy position of the Mn d-states (related to the peak A in Figure 4a) that strongly depends on the core-hole Coulomb potential.

### 4.3 Orientation of Mn magnetic moments

In order to study a magnetic order in the material we have applied the x-ray magnetic circular dichroism technique. While XES and spin-polarized XANES are sensitive to the absolute value of spin magnetic moment, the XMCD intensity depends on the magnetic moment projection on the global axis set by external magnetic field[60, 61]. Figure 5a shows the circular dichroism signal in Mn $L_{2,3}$ XANES measured at 10 K temperature and 5 T external magnetic field. Labels $\mu^+$ and $\mu^-$ stand for the directions of the photon helicity. Positive direction of the applied magnetic field is collinear with the photon propagation direction.

Spin and orbital components of the shell-specific magnetic moment ($m_{spin}$, $m_{orb}$) can be estimated using sum rules for the single ion[60-62]. The degree of polarization of incident X-rays and the saturation of XMCD signal with the applied magnetic field should be taken into account. In the experiment, the Stokes parameter S3 was equal to 0.9. The XMCD signal was measured for the B values from 0T to 5T with step 0.2T and the saturated value was obtained by fitting with a Brillouin function (see

Figures S7-S9 in supplementary information and discussion about anisotropic behaviour of magnetization in ZnO:Mn). We obtain $m_{spin}=0.39n_h$ where $n_h$ stands for the number of holes in the Mn d-shell. The nonzero orbital magnetic moment was found with a value of $0.02n_h$. Unfortunately values obtained from the sum rules for $3d^5$ system suffer from large errors[63] since 2p-3d Coulomb interaction leads to the mixing of the j=3/2 and j=1/2 states. Thus these values cannot be directly compared with the Mn spin magnetic moment obtained from XES analysis in the previous section. Instead one has to use correction factors to determine $m_{spin}$ value[64] or compare the results of the sum rules analysis for experimental and simulated spectra. Below we discuss the difference between Mn spin magnetic moments derived from experimental XMCD and theoretically calculated ones for the $Mn^{2+}$ ion in $3d^5$ configuration.

The atomic charge state is the main input parameter of the atomic multiplet simulations. The charge state of Mn was determined from Bader analysis[65, 66] of the electron density obtained for ZnO:Mn and Mn oxides by means of GGA PBE simulations within the FLAPW approximation. The results are listed in the Table 1. The Bader charges for Mn in MnO and $Mn_{Zn}$ are similar (+1.35 and +1.37 correspondingly). Thus the formal charge state of $Mn_{Zn}$ is 2+. We also found that calculated charge for the four oxygens around Mn is larger than in pure ZnO. This can be explained by the hybridization between Mn d-states and O p-states, which should be considered when applying XMCD sum rules. For a single $Mn^{2+}$ ion the number of holes is 5, but for $Mn^{2+}$ in ZnO the hybridization of the Mn d-states with O p-states reduces the number of Mn-localized d-holes. As follows from the integration of the $Mn_{Zn}$ d-DOS obtained from FLAPW simulations, $n_h$ is approximately 4.2.

Atomic multiplet simulations are provided in Figure 5b for the $Mn^{2+}$ ion with $3d^5$ configuration. 10Dq parameter was chosen equal to -0.5 and satisfactory agreement with experiment was obtained even without taking into account charge transfer effects that influence mainly the energy position of peak C in the Mn $L_{2,3}$ spectra. Applying sum rules to the calculated spectra we obtain $m_{orb}=0$ and $m_{spin}=0.64n_h$. Therefore if all Mn moments had been aligned collinearly, the sum rules applied to the experimental Mn $L_{2,3}$ XMCD spectra would have also produced $m_{spin}=0.64n_h$. However, the value of $m_{spin}=0.39n_h$ was obtained in the experiment. Thus one can conclude that only 60±5% of all manganese moments form paramagnetic phase being aligned coherently by external magnetic field. According to EXAFS data analysis, 10±5% of Mn atoms can form secondary phases which means that at least 30±10% of Mn atoms substitute Zn sites and are coupled antiferromagnetically. The antiferromagnetic order of neighbouring Mn moments in dilute ZnO:Mn was also predicted theoretically[67, 68].

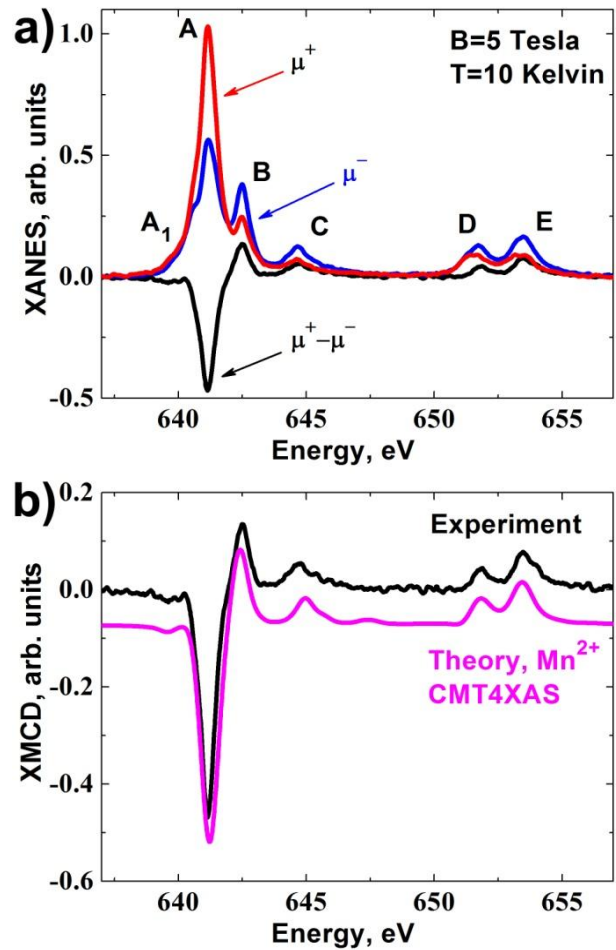

**Figure 5**. (a) Mn $L_{2,3}$ XANES spectra, measured for different photon helicities for ZnO:Mn and corresponding XMCD signal. (b) Experimental XMCD signal compared to the atomic multiplet simulations.

**Table 1**. Bader analysis of the charge states of atoms in pure ZnO, MnO and in ZnO:Mn. In case of ZnO:Mn the average value for the four oxygen atoms in the $MnO_4$ tetrahedron is presented (tolerance = 0.01e, GGA PBE calculation).

|    | ZnO   | ZnO:Mn | MnO   | $Mn_3O_4$ | $Mn_2O_3$ | $MnO_2$ |
|----|-------|--------|-------|-----------|-----------|---------|
| Zn | +1.22 | +1.22  |       |           |           |         |
| O  | -1.22 | -1.25  | -1.35 | -1.20     | -1.14     | -0.94   |
| Mn |       | +1.37  | +1.35 | +1.60     | +1.71     | +1.88   |

## 5. Conclusions

EXAFS data analysis revealed that a shell of the ZnO/ZnO:Mn core-shell nanoneedles consisted of dilute ZnO:Mn and 90%±5% of Mn atoms substituted Zn. The spectrum above Mn K-edge showed XLD while pre-edge peak was isotropic. The origin of such a behaviour is that ZnO:Mn shell has a wurtzite crystalline structure with a growth axis parallel to the ZnO nanowires, while pre-edge electron states are localized mainly within first coordination sphere of Mn with Td symmetry. Spin-polarized Mn K-edge XANES were measured in the HERFD mode by setting fluorescence detection energies to the maxima of $K\beta_{1,3}$ and $K\beta'$ emission lines, *i.e.* 6491.8 eV and 6476.2 eV respectively. It was experimentally observed that Mn p-projected DOS, observed in XANES, are partially spin-polarized, while pre-edge electron states (Mn-d and O-p hybridized) are totally spin-polarized relatively to the local orientation of the d-shell magnetic moment of absorbing atom. The XES data complemented by theoretical Bader analysis showed that Mn magnetic moment as well as its charge state are close to ones in MnO. The value of Mn magnetic moment was estimated to be $4.3\mu_B$. The orientation of Mn magnetic moments in the applied external field was analyzed by XMCD. By applying spin sum rule we found a smaller value of Mn spin magnetic moment than theoretically predicted for $Mn^{2+}$ ion, which imply the coexistence of paramagnetic and antiferromagnetic phases in the dilute ZnO:Mn shell.

## Acknowledgements

The research was supported by the Russian Ministry of Education and Science (RNP 2.1.1.5932). We acknowledge the European Synchrotron Radiation Facility and Helmholtz-Zentrum Berlin, BESSY II, for provision of synchrotron radiation. N.S. and A.G. would like to thank Russian Ministry of Education for providing the fellowships of President of Russian Federation to study abroad. We also would like to thank the UGINFO computer center of Southern Federal University for providing the computer time and Dr. V.N. Datsyuk for assistance.

## Notes and references

† Electronic Supplementary Information (ESI) containing details about theoretical simulations and XMCD data analysis is available: See DOI: 10.1039/b000000x/

1. F. Pan, C. Song, X. J. Liu, Y. C. Yang and F. Zeng, *Materials Science and Engineering: R: Reports*, 2008, **62**, 1-35. DOI: 10.1016/j.mser.2008.04.002.
2. Y. C. Yang, F. Pan, Q. Liu, M. Liu and F. Zeng, *Nano Letters*, 2009, **9**, 1636-1643. DOI: 10.1021/nl900006g.
3. T. Dietl, H. Ohno, F. Matsukura, J. Cibert and D. Ferrand, *Science*, 2000, **287**, 1019-1022. DOI: 10.1126/science.287.5455.1019.
4. P. Sharma, A. Gupta, K. V. Rao, F. J. Owens, R. Sharma, R. Ahuja, J. M. O. Guillen, B. Johansson and G. A. Gehring, *Nat Mater*, 2003, **2**, 673-677. DOI: http://www.nature.com/nmat/journal/v2/n10/suppinfo/nmat984_S1.html.
5. Z. Yang, J. L. Liu, M. Biasini and W. P. Beyermann, *Applied Physics Letters*, 2008, **92**, 042111-042113.
6. D. F. Wang, S. Y. Park, Y. S. Lee, Y. P. Lee, J. C. Li and C. Liu, *Journal of Applied Physics*, 2008, **103**, 07D126-123.
7. E. Céspedes, J. Sánchez-Marcos, J. García-López and C. Prieto, *Journal of Magnetism and Magnetic Materials*, 2010, **322**, 1201-1204. DOI: http://dx.doi.org/10.1016/j.jmmm.2009.05.038.
8. M. Venkatesan, C. B. Fitzgerald, J. G. Lunney and J. M. D. Coey, *Physical Review Letters*, 2004, **93**, 177206.
9. Z. Linjuan, L. Jiong, D. Yaping, W. Jianqiang, W. Xiangjun, Z. Jing, C. Jie, C. Wangsheng, J. Zheng, H. Yuying, Y. Chunhua, Z. Shuo and W. Ziyu, *New Journal of Physics*, 2012, **14**, 013033.
10. M. Naeem and S. K. Hasanain, *Journal of Physics: Condensed Matter*, 2012, **24**, 245305.
11. Q. Wang, Q. Sun, G. Chen, Y. Kawazoe and P. Jena, *Physical Review B*, 2008, **77**, 205411.
12. N. H. Hong, J. Sakai and V. Brizé, *Journal of Physics: Condensed Matter*, 2007, **19**, 036219.
13. B. B. Straumal, A. A. Mazilkin, S. G. Protasova, A. A. Myatiev, P. B. Straumal, G. Schütz, P. A. van Aken, E. Goering and B. Baretzky, *Physical Review B*, 2009, **79**, 205206.
14. A. Ney, M. Opel, T. C. Kaspar, V. Ney, S. Ye, K. Ollefs, T. Kammermeier, S. Bauer, K. W. Nielsen, S. T. B. Goennenwein, M. H. Engelhard, S. Zhou, K. Potzger, J. Simon, W. Mader, S. M. Heald, J. C. Cezar, F. Wilhelm, A. Rogalev, R. Gross and S. A. Chambers, *New Journal of Physics*, 2010, **12**, 013020.
15. R. Lardé, E. Talbot, F. Pareige, H. Bieber, G. Schmerber, S. Colis, V. Pierron-Bohnes and A. Dinia, *Journal of the American Chemical Society*, 2011, **133**, 1451-1458. DOI: 10.1021/ja108290u.
16. C. Sudakar, J. S. Thakur, G. Lawes, R. Naik and V. M. Naik, *Physical Review B*, 2007, **75**, 054423.
17. A. Ney, K. Ollefs, S. Ye, T. Kammermeier, V. Ney, T. C. Kaspar, S. A. Chambers, F. Wilhelm and A. Rogalev, *Physical Review Letters*, 2008, **100**, 157201.
18. F. Schoofs, T. Fix, A. M. H. R. Hakimi, S. S. Dhesi, G. van der Laan, S. A. Cavill, S. Langridge, J. L. MacManus-Driscoll and M. G. Blamire, *Journal of Applied Physics*, 2010, **108**, 053911-053915.
19. A. Ney, V. Ney, F. Wilhelm, A. Rogalev and K. Usadel, *Physical Review B*, 2012, **85**, 245202.
20. T. Kataoka, Y. Yamazaki, V. R. Singh, Y. Sakamoto, A. Fujimori, Y. Takeda, T. Ohkochi, S. I. Fujimori, T. Okane, Y. Saitoh, H. Yamagami, A. Tanaka, M. Kapilashrami, L. Belova and K. V. Rao, *Applied Physics Letters*, 2011, **99**, 132508-132503.
21. K. Hämäläinen, C. C. Kao, J. B. Hastings, D. P. Siddons, L. E. Berman, V. Stojanoff and S. P. Cramer, *Physical Review B*, 1992, **46**, 14274-14277.
22. G. Peng, F. M. F. deGroot, K. Haemaelaeinen, J. A. Moore, X. Wang, M. M. Grush, J. B. Hastings, D. P. Siddons and W. H. Armstrong, *Journal of the American Chemical Society*, 1994, **116**, 2914-2920. DOI: 10.1021/ja00086a024.
23. G. D. Pirngruber, J.-D. Grunwaldt, J. A. van Bokhoven, A. Kalytta, A. Reller, O. V. Safonova and P. Glatzel, *The Journal of Physical Chemistry B*, 2006, **110**, 18104-18107. DOI: 10.1021/jp063812b.
24. M. Lorenz, E. M. Kaidashev, A. Rahm, T. Nobis, J. Lenzner, G. Wagner, D. Spemann, H. Hochmuth and M. Grundmann, *Applied Physics Letters*, 2005, **86**, 143113-143113.
25. V. E. Kaydashev, E. M. Kaidashev, M. Peres, T. Monteiro, M. R. Correia, N. A. Sobolev, L. C. Alves, N. Franco and E. Alves, *Journal of Applied Physics*, 2009, **106**, 093501-093504.
26. P. Glatzel and U. Bergmann, *Coordination Chemistry Reviews*, 2005, **249**, 65-95. DOI: 10.1016/j.ccr.2004.04.011.
27. C. Gauthier, V. A. Sole, R. Signorato, J. Goulon and E. Moguiline, *Journal of Synchrotron Radiation*, 1999, **6**, 164-166. DOI: doi:10.1107/S0909049598016835.
28. P. Blaha, K. Schwarz, G. K. H. Madsen, D. Kvasnicka and J. Luitz, 2001.
29. P. Blaha, K. Schwarz, P. Sorantin and S. B. Trickey, *Computer Physics Communications*, 1990, **59**, 399-415. DOI: 10.1016/0010-4655(90)90187-6.
30. J. P. Perdew, K. Burke and M. Ernzerhof, *Physical Review Letters*, 1996, **77**, 3865-3868.
31. U. Ozgur, Y. I. Alivov, C. Liu, A. Teke, M. A. Reshchikov, S. Dogan, V. Avrutin, S. J. Cho and H. Morkoc, *Journal of Applied Physics*, 2005, **98**, 041301-041103.
32. E. Stavitski and F. M. F. de Groot, *Micron*, 2010, **41**, 687-694. DOI: 10.1016/j.micron.2010.06.005.
33. A. A. Guda, N. Smolentsev, J. Verbeeck, E. M. Kaidashev, Y. Zubavichus, A. N. Kravtsova, O. E. Polozhentsev and A. V. Soldatov, *Solid State Communications*, 2011, **151**, 1314-1317. DOI: 10.1016/j.ssc.2011.06.028.
34. A. Rahm, E. M. Kaidashev, H. Schmidt, M. Diaconu, A. Pöppl, R. Böttcher, C. Meinecke, T. Butz, M. Lorenz and M. Grundmann, *Microchimica Acta*, 2006, **156**, 21-25. DOI: 10.1007/s00604-006-0602-1.
35. S. Zhou, K. Potzger, Q. Xu, G. Talut, M. Lorenz, W. Skorupa, M. Helm, J. Fassbender, M. Grundmann and H. Schmidt, *Vacuum*, 2009, **83, Supplement 1**, S13-S19. DOI: 10.1016/j.vacuum.2009.01.030.
36. J. J. Rehr, J. J. Kas, F. D. Vila, M. P. Prange and K. Jorissen, *Physical Chemistry Chemical Physics*, 2010, **12**, 5503-5513.
37. T. Devillers, M. Rovezzi, N. G. Szwacki, S. Dobkowska, W. Stefanowicz, D. Sztenkiel, A. Grois, J. Suffczyński, A. Navarro-Quezada, B. Faina, T. Li, P. Glatzel, F. d'Acapito, R. Jakieła, M. Sawicki, J. A. Majewski, T. Dietl and A. Bonanni, *Sci. Rep.*, 2012, **2**. DOI: http://www.nature.com/srep/2012/121010/srep00722/abs/srep00722.html#supplementary-information.
38. N. Smolentsev, A. V. Soldatov, G. Smolentsev and S. Q. Wei, *Solid State Communications*, 2009, **149**, 1803-1806. DOI: http://dx.doi.org/10.1016/j.ssc.2009.07.016.
39. M. C. Mugumaoderha, R. Sporken, J. Ghijsen, F. M. F. de Groot and J. A. Dumont, *The Journal of Physical Chemistry C*, 2011, **116**, 665-670. DOI: 10.1021/jp206705p.
40. C. Guglieri, E. Céspedes, C. Prieto and J. Chaboy, *Journal of Physics: Condensed Matter*, 2011, **23**, 206006.

41. G. Ciatto, A. Di Trolio, E. Fonda, P. Alippi, A. M. Testa and A. A. Bonapasta, *Physical Review Letters*, 2011, **107**, 127206.
42. C. Brouder, *Journal of Physics: Condensed Matter*, 1990, **2**, 701.
43. B. Gilbert, B. H. Frazer, H. Zhang, F. Huang, J. F. Banfield, D. Haskel, J. C. Lang, G. Srajer and G. D. Stasio, *Physical Review B*, 2002, **66**, 245205.
44. L. Xue-Chao, S. Er-Wei, C. Zhi-Zhan, C. Bo-Yuan, Z. Tao, S. Li-Xin, Z. Ke-Jin, C. Ming-Qi, Y. Wen-Sheng, X. Zhi, H. Bo and W. Shi-Qiang, *Journal of Physics: Condensed Matter*, 2008, **20**, 025208.
45. W. Yan, Z. Sun, Q. Liu, Z. Li, Z. Pan, J. Wang, S. Wei, D. Wang, Y. Zhou and X. Zhang, *Applied Physics Letters*, 2007, **91**, 062113-062113.
46. T. Isao and O. Fumiyasu, *Journal of Physics: Condensed Matter*, 2008, **20**, 064215.
47. K. Tsutsumi, H. Nakamori and K. Ichikawa, *Physical Review B*, 1976, **13**, 929-933.
48. G. Vankó, T. Neisius, G. Molnár, F. Renz, S. Kárpáti, A. Shukla and F. M. F. de Groot, *The Journal of Physical Chemistry B*, 2006, **110**, 11647-11653. DOI: 10.1021/jp0615961.
49. G. Vankó, J.-P. Rueff, A. Mattila, Z. Németh and A. Shukla, *Physical Review B*, 2006, **73**, 024424.
50. S. Limandri, S. Ceppi, G. Tirao, G. Stutz, C. G. Sánchez and J. A. Riveros, *Chemical Physics*, 2010, **367**, 93-98. DOI: 10.1016/j.chemphys.2009.11.001.
51. S. A. Pizarro, P. Glatzel, H. Visser, J. H. Robblee, G. Christou, U. Bergmann and V. K. Yachandra, *Physical Chemistry Chemical Physics*, 2004, **6**, 4864-4870.
52. X. Wang, F. M. F. de Groot and S. P. Cramer, *Physical Review B*, 1997, **56**, 4553-4564.
53. F. M. F. de Groot, S. Pizzini, A. Fontaine, K. Hämäläinen, C. C. Kao and J. B. Hastings, *Physical Review B*, 1995, **51**, 1045-1052.
54. P. Thakur, K. H. Chae, J. Y. Kim, M. Subramanian, R. Jayavel and K. Asokan, *Applied Physics Letters*, 2007, **91**, 162503-162503.
55. V. I. Anisimov, I. V. Solovyev, M. A. Korotin, M. T. Czyżyk and G. A. Sawatzky, *Physical Review B*, 1993, **48**, 16929-16934.
56. G. Madsen, K. H. and P. Novák, *Europhys. Lett.*, 2005, **69**, 777-783.
57. M. Toyoda, H. Akai, K. Sato and H. Katayama-Yoshida, *Physica B: Condensed Matter*, 2006, **376–377**, 647-650. DOI: http://dx.doi.org/10.1016/j.physb.2005.12.163.
58. G. C. Zhou, L. Z. Sun, X. L. Zhong, X. Chen, L. Wei and J. B. Wang, *Physics Letters A*, 2007, **368**, 112-116. DOI: 10.1016/j.physleta.2007.03.061.
59. V. I. Anisimov, F. Aryasetiawan and A. I. Lichtenstein, *Journal of Physics: Condensed Matter*, 1997, **9**, 767.
60. P. Carra, B. T. Thole, M. Altarelli and X. Wang, *Physical Review Letters*, 1993, **70**, 694-697.
61. B. T. Thole, P. Carra, F. Sette and G. van der Laan, *Physical Review Letters*, 1992, **68**, 1943-1946.
62. C. T. Chen, Y. U. Idzerda, H. J. Lin, N. V. Smith, G. Meigs, E. Chaban, G. H. Ho, E. Pellegrin and F. Sette, *Physical Review Letters*, 1995, **75**, 152-155.
63. C. Piamonteze, P. Miedema and F. M. F. de Groot, *Physical Review B*, 2009, **80**, 184410.
64. K. W. Edmonds, N. R. S. Farley, T. K. Johal, G. van der Laan, R. P. Campion, B. L. Gallagher and C. T. Foxon, *Physical Review B*, 2005, **71**, 064418.
65. R. F. W. Bader, *Clarendon Press, Oxford, UK*, 1990.
66. R. F. W. Bader, *Chemical Reviews*, 1991, **91**, 893-928. DOI: 10.1021/cr00005a013.
67. M. H. F. Sluiter, Y. Kawazoe, P. Sharma, A. Inoue, A. R. Raju, C. Rout and U. V. Waghmare, *Physical Review Letters*, 2005, **94**, 187204.
68. T. Chanier, M. Sargolzaei, I. Opahle, R. Hayn and K. Koepernik, *Physical Review B*, 2006, **73**, 134418.